\documentclass[aps,prl,10pt,twocolumn,floatfix,showpacs,superscriptaddress,longbibliography]{revtex4-1}

\usepackage{graphicx,amsmath,amsfonts,bm}
\usepackage[colorlinks=true, citecolor=blue, linkcolor=blue, urlcolor=blue]{hyperref}
\usepackage[all]{hypcap}

\begin{document}

\title{Correlation and current anomalies in helical quantum dots}
\author{C. De Beule}
\email{christophe.debeule@uantwerpen.be}
\affiliation{Department of Physics, University of Antwerp, 2020 Antwerp, Belgium}
\author{N. Traverso Ziani}
\email{niccolo.traverso@physik.uni-wuerzburg.de}
\affiliation{Institute of Theoretical Physics and Astrophysics, University of W\"urzburg, 97074 W\"urzburg, Germany}
\author{M. Zarenia}
\affiliation{Department of Physics, University of Antwerp, 2020 Antwerp, Belgium}
\author{B. Partoens}
\affiliation{Department of Physics, University of Antwerp, 2020 Antwerp, Belgium}
\author{B. Trauzettel}
\affiliation{Institute of Theoretical Physics and Astrophysics, University of W\"urzburg, 97074 W\"urzburg, Germany}
\affiliation{Department of Physics, University of California, Berkeley, California 94720, USA}
\begin{abstract}
We theoretically investigate the ground-state properties of a quantum dot defined on the surface of a strong three-dimensional time-reversal invariant topological insulator. Confinement is realized by ferromagnetic barriers and Coulomb interaction is treated numerically for up to seven electrons in the dot. Experimentally relevant intermediate interaction strengths are considered. The topological origin of the dot has interesting consequences: $i)$ spin polarization increases and the ground state exhibits quantum phase transitions at specific angular momenta as a function of interaction strength $ii)$ the onset of Wigner correlations takes place mainly in one spin channel, $iii)$ the ground state is characterized by a persistent current that changes sign as a function of the radius of the dot.
\end{abstract}

\maketitle

Topological insulators (TIs) \cite{2dteo1,2dteo2,2dexp1,2dexp2,3dteo1,3dteo2,3dexp} are peculiar insulators hosting metallic states at boundaries with other insulators distinguished by their topology.
The metallic surface or interface states are, for a certain class of TIs, protected from time-reversal invariant perturbations and characterized by spin-momentum locking, relating the quasi-momentum of the one-particle states to their spin projection \cite{rev1,rev2}. Due to their special properties, the surface states are potentially interesting for applications in spintronics \cite{st1,st2,st3} and quantum computation \cite{qc1,qc2}. A promising step toward device applications is represented by the progress made in topological insulator nanostructures. On the one hand, nanostructures can overcome some intrinsic limitations of state of the art TIs \cite{short,nanostruct1,nanostruct2,nanostruct3}. On the other hand, the ability to produce topological nanostructures may lead to new physics, relating to, for instance, Majorana fermions \cite{majo1,majo2,majo3}, parafermions \cite{para1,para2,para3}, qubits \cite{qbit}, topological Josephson junctions \cite{jj1,jj2,jj3}, and 
anomalous superconducting pairing \cite{francois}.\\
\indent The main challenge in the theoretical description of nanostructured TIs is represented by the need to properly take into account electron-electron interactions. Interaction effects in two-dimensional quantum spin Hall systems are striking: In the weakly interacting regime, the low-energy properties of the edge states are well described by the helical Luttinger liquid \cite{helical1,helical2}; in the case of strong interactions, the physics is even richer, in terms of both exotic edge \cite{para1,para2,fw} and bulk \cite{fract} properties. For three-dimensional time-reversal invariant TIs, the Landau theory of the two-dimensional helical surface states has been developed to deal with the weak to intermediate interaction regime \cite{helland}, while the strongly interacting regime is dominated by fractionalization \cite{fract}.\\
\indent Moreover, the study of interacting massless Dirac fermions, relevant for strong TIs, can greatly benefit from research on the electronic properties of graphene \cite{graphene}. Graphene has been deeply investigated during the last decade and it was indeed shown that the properties of interacting particles in nanostructures can significantly differ from the corresponding bulk theories. For example, bulk graphene is characterized by the absence of Wigner crystallization \cite{nowig}: due to the linear spectrum, the kinetic energy has the same scaling with density as the Coulomb energy 
so that there is no competition.
Consequently, intrinsic interactions in graphene are insufficient to establish Wigner crystallization. In finite structures, such as quantum dots, this simple picture fails, and Wigner molecules, or their onset, may emerge for strong but reasonable interaction strengths \cite{egger,karina}. A similar increase of the effects of electron-electron interactions in finite structures has been reported \cite{egger2} for quantum dots in a two-dimensional electron gas (2DEG) \cite{2d1,2d2,2d3,2d4}. In general, the same behavior also characterizes the surface states of strong three-dimensional TIs. However, as we show below, the presence of a \emph{single} Dirac cone has remarkable consequences on the ground state of topological quantum dots.\\
\indent In this Letter, we investigate the ground-state properties of a circular quantum dot hosted on the surface of a three-dimensional strong topological insulator, confined by hard wall (i.e.\ infinite mass) boundaries. The inspection is carried out numerically with exact diagonalization for up to seven electrons in the dot. We consider an intermediate interaction regime such that electrons are strongly correlated but the Wigner molecule is not yet fully developed, meaning that the electrons are not yet fully localized. After introducing the model, we discuss the anomalous spin-resolved behavior of the ground state. The quantum dot is characterized by a spontaneous spin polarization which increases with interaction strength, similar to quantum dots in a 2DEG with strong Rashba coupling \cite{naseri}. We draw an analogy, in terms of quantum phase transitions, between interactions in the topological system and external magnetic fields in 2DEG quantum dots. Moreover, the spatial distribution of the spin-resolved density shows a partial segregation between spins. While the density of the majority spin (spin up in our case) is localized mostly near the boundary of the dot, that of the minority spin (spin down in our case) is localized more near the center. Intuitively, this suggests that spin down particles are effectively stiffer towards electron-electron interactions and, hence, energetically unfavored. This picture is supported by the spin-resolved density-density correlation functions: the majority spin component shows strong correlations, while the minority spin density remains more liquid-like. Finally, we present the most remarkable property of the topological dot: The ground state is characterized by a persistent current. Surprisingly, the current changes sign as a function of the radius of the dot in all cases under consideration. We carefully explain below why this happens.

The Hamiltonian $H$ we consider is
\begin{equation}
H=\sum_i H_0^{(i)}+V_C, \quad H_0^{(i)}=T^{(i)}+U^{(i)}.
\end{equation}
Here, $H_0$ is the one-particle Hamiltonian and $V_C$ the Coulomb repulsion. For the one-particle wave functions, we adopt the spinor basis $\psi(\bm{r}^{(i)})=\left( \psi_+(\bm{r}^{(i)}),-i\psi_-(\bm{r}^{(i)}) \right)^T$, where the coordinate $\bm{r}^{(i)}=(x^{(i)},y^{(i)})$ represents the position of the $i$-th electron on the TI surface. In this basis, the kinetic term $T^{(i)}$ reads $T^{(i)}=v_F\bm{\sigma}^{(i)}\cdot\bm{p}^{(i)}$, where $v_F$ is the Fermi velocity, $\bm{\sigma}^{(i)}=(\sigma_x^{(i)},\sigma_y^{(i)})$ are the Pauli matrices acting on the spin of the $i$-th particle, and $\bm{p}^{(i)}=-i\hbar\nabla_{\bm r^{(i)}}$. The confining potential $U^{(i)}$ we consider is $U^{(i)}=M \Theta(r^{(i)}-R)\sigma_z^{(i)}$, where $r^{(i)}=|\bm{r}^{(i)}|$, $\Theta(x)$ is the Heaviside step function, $R$ is the radius of the dot, and the limit $M\rightarrow \infty$ is taken for simplicity \cite{berry}. Note that this confinement breaks time-reversal symmetry.
Experimentally, such confinement could be realized by depositing a structured ferromagnetic layer on the TI surface. The (spinor) eigenvectors $\psi_{n,m}(\bm{r}^{(i)})$ of $H_0^{(i)}$ are
\begin{equation}
\psi_{n,m}(\bm{r})= A_{n,m} e^{im\theta} \left(\begin{matrix} J_m (k_{n,m}r) \\ ie^{i\theta} J_{m+1} (k_{n,m}r) \end{matrix}\right).
\end{equation}
Here, $m$ is an integer, $\theta=\arctan\left( y/x \right)$, $J_m(x)$ are Bessel functions, and $k_{n,m}$ are the solutions for $k$ with $n$ radial nodes of the hard wall boundary condition:
\begin{equation}
J_m(kR)=J_{m+1}(kR).
\end{equation}
The spinor normalization then reads $A_{n,m}^{-2} = 2\pi R^2 J_m(k_{n,m}R)^2 \left[ 2-(2m+1)/(k_{n,m}R) \right]$ and the corresponding eigenvalues are $E_{n,m}=\hbar v_F k_{n,m}$. The spectrum is symmetric with respect to zero energy and a gap $\Delta \approx 2.87\hbar v_F/R$ is present. Throughout this Letter, similarly to Refs.\ \cite{egger,plus1,plus2,plus3}, we assume all negative energy states to be filled and inert. For $(n,m)$ corresponding to positive energy, we define the fermionic annihilation (creation) operators $c_{n,m}$ ($c^\dag_{n,m}$). The corresponding Fermi spinor $\Psi(\bm{r})$ is given by $\Psi(\bm{r})=(\Psi_+(\bm{r}),\Psi_-(\bm{r}))^T=\sum_{n,m}\psi_{n,m}(\bm{r})c_{n,m}$. The explicit form of the Hamiltonian, once the negative energy states are removed from the Hilbert space, is (repeated indexes are summed)
\begin{eqnarray}
H'&=&H_0'+V'_C, \label{eq:H} \\
H'_0&=&E_{n,m} c^\dag_{n,m}c_{n,m},\\
V'_C&=&V_{n_1,m_1;n_2,m_2}^{n_3,m_3;n_4,m_4}c^\dag_{n_1,m_1}c^\dag_{n_2,m_2}c_{n_3,m_3}c_{n_4,m_4},
\end{eqnarray}
where $V_{n_1,m_1;n_2,m_2}^{n_3,m_3;n_4,m_4}$ are the matrix elements of the Coulomb interaction with respect to the spinor wave functions $\psi_{n_i,m_i}(\bm{r}^{(i)})$. Since the total angular momentum $J_z = \sum_i \left( m_i + 1/2 \right)$ is conserved in our model, the Hamiltonian can be written in block diagonal form.
The effective fine-structure constant $\alpha$ that controls the relative weight of kinetic and Coulomb energy is given by $\alpha=e^2/(4\pi \epsilon_0\epsilon \hbar v_F)$, where 
$\epsilon$ is the surface dielectric constant of the TI under inspection. As expected, $\alpha$ is independent of the density, and depends only on $v_F$ and $\epsilon$.
For $\rm{Bi}_2\rm{Te}_3$ thin films \cite{fermivel} on $\rm{SiO}_2$ \cite{sio2}, one finds $\alpha \approx 1.8$, which could increase further by strain or substrate engineering \cite{nanostructuring} to reduce $v_F$ and $\epsilon$, respectively. Moreover, strongly interacting electrons are to be expected on the surfaces of TIs with a strongly interacting bulk such as Heusler compounds, whose bulk electrons show heavy fermion behavior \cite{heusler1}.

\begin{figure}
	\centering
	\includegraphics[width=\linewidth]{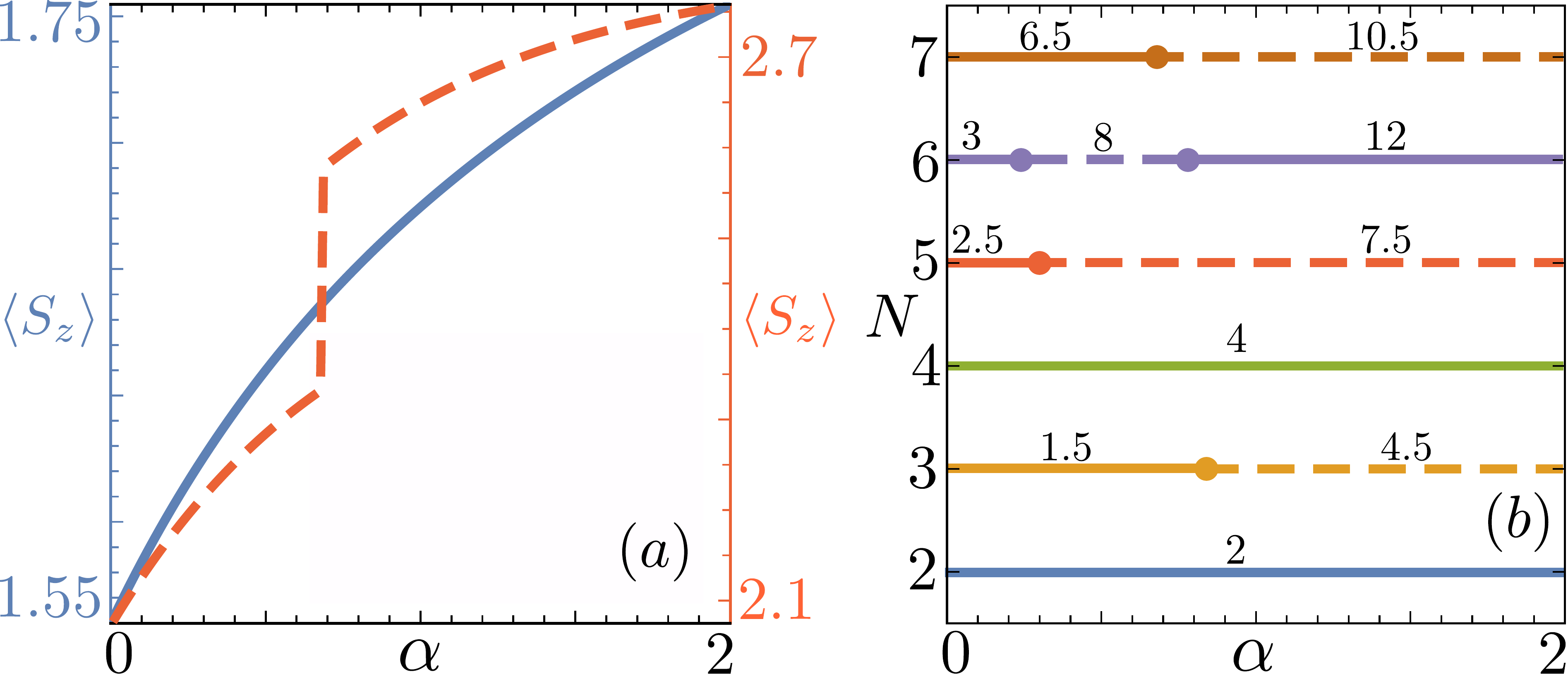}
	\caption{(Color online)$(a)$ Spin polarization $\left<S_z\right>$, in units $\frac{\hbar}{2}$, as a function of $\alpha$ for  $N=4$ (blue, solid, left $y$-axis), and $N=7$ (red, dashed, right $y$-axis). $(b)$ Quantum phase diagram as a function of $N$ and $\alpha$. Dots indicate phase transitions and the numbers are the angular momentum $J_z$ of the ground state.}
	\label{fig1}
\end{figure}

Our aim is to characterize the ground-state properties of a dot containing $N$ electrons, described by the Hamiltonian in Eq.\ \eqref{eq:H} in the intermediate interaction regime  $0<\alpha<2$. To do so, the ground-state average $\langle O \rangle$ of operators $O$ are evaluated numerically by exact diagonalization. The first quantity we address is the average spin polarization $\left<S_z\right> = \frac{\hbar}{2} \int d^2\bm r \langle\Psi^\dag(\bm{r})\sigma_z\Psi(\bm{r})\rangle$. The noninteracting spin polarization $\left<S_z\right>$ is due to the presence of the boundary that breaks time-reversal symmetry. Turning on electron-electron interactions increases the spin polarization: two different cases are shown in Fig.\ \ref{fig1}(a). For $N=4$ the increase is smooth, while for $N=7$ the increase is larger and shows a jump. As shown in Fig.\ \ref{fig1}(b), the difference is due to the nature of the ground state as a function of $\alpha$: while for $N=4$ one has total angular momentum $J_z=4$ for $0<\alpha<2$, the jump in $\left<S_z\right>$ for $N=7$ corresponds to a level crossing where $J_z$ changes from $J_z=6.5$ to $J_z=10.5$ at $\alpha \approx 0.7$. The physical significance of the polarization is the following: With increasing interaction strength, the system evolves towards a classical state, the Wigner molecule. In non-topological systems, the classical nature, and the consequent loss of the spin degree of freedom, is allowed by the fact that many states tend to become degenerate in the limit of infinite interaction strength. The combination of a single cone and spin-momentum locking prevents our system from developing this degeneracy. Increasing the spin polarization is therefore a natural way to approach the classical limit in the topological quantum dot. The quantum phase diagram shown in Fig.\ \ref{fig1}(b) indicates that the effect is generic for $2<N<7$ and $0<\alpha<2$. Moreover, it presents some peculiarities: in ordinary 2DEG quantum dots, $\left< S_z \right>$ does not change with interaction strength. In this case the total spin is fixed by an applied magnetic field which itself can induce quantum phase transitions towards states with a higher spin imbalance \cite{egger2,2d1}. The \emph{magic} numbers, i.e.\ the orbital angular momentum $M$ of the ground state, can then be determined from the symmetry properties of the wave function (e.g.\ $M=3, 6, \ldots$ for $N=3$ and $M=6, 10, \ldots$ for $N=4$) \cite{newref}. The interplay of spin-momentum locking, Coulomb interaction, and the boundary which act as a spin-selection mechanism yields similar (but different) quantum phase transitions in our system at zero magnetic field as $\alpha$ is increased. Most strikingly, the analogy is profound: the observed $J_z$ ground state values can be obtained from assuming a maximum value for $\left< S_z \right>$, and the same \emph{magic} number rule for the angular momentum expectation value, even if the maximal spin value is not yet reached.

\begin{figure}
	\centering
	\includegraphics[width=\linewidth]{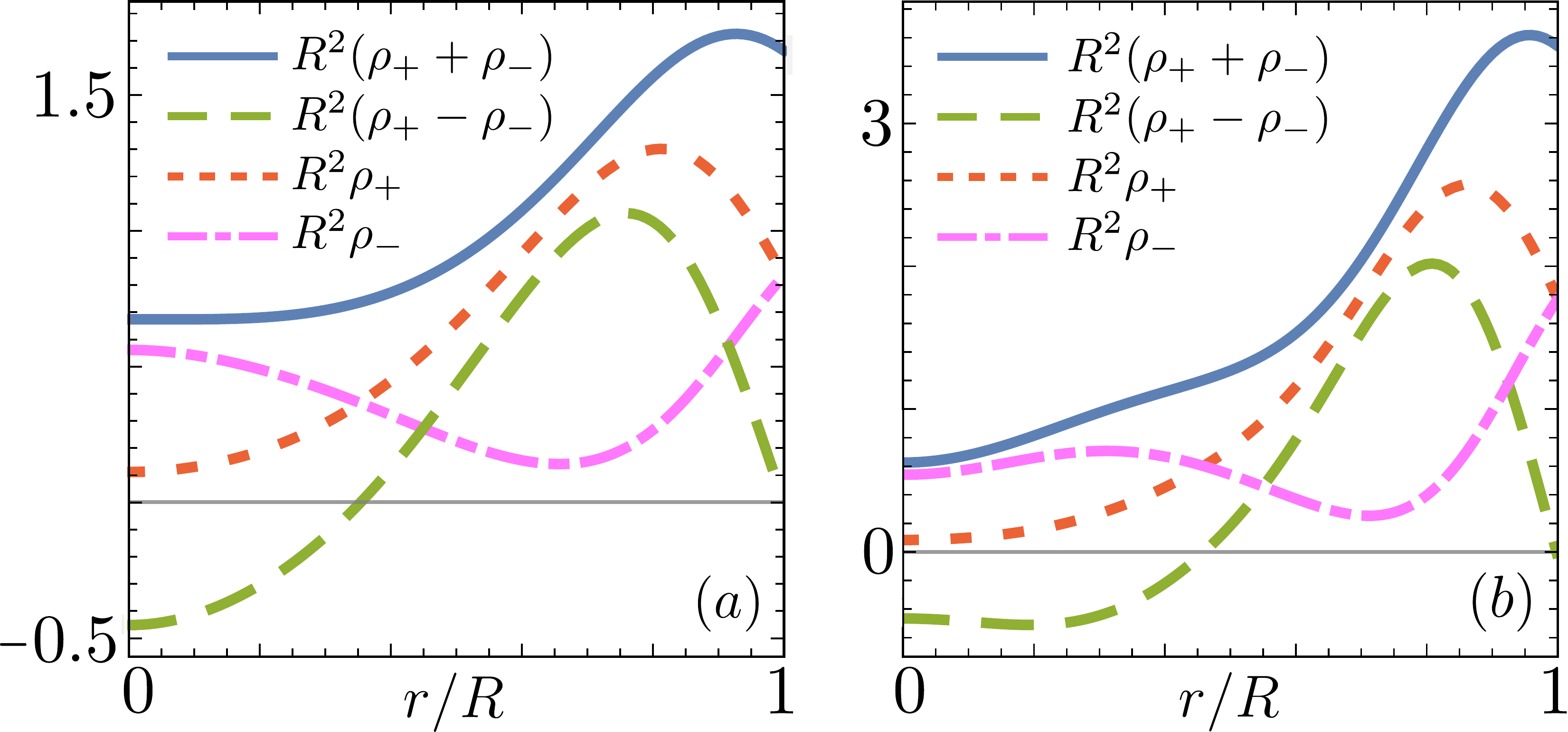}
	\caption{(Color online) Total electron density $\rho_{+}(r)+\rho_{-}(r)$ (red, solid), spin imbalance $\rho_{+}(r)-\rho_{-}(r)$ (blue, dashed), $\rho_{+}(r)$ (green, dotted), and $\rho_{-}(r)$ (purple, dash dotted) as a function of the radius $r$ for $\alpha=2$ and $(a)$ $N=4$, $(b)$ $N=7$.}
	\label{fig2}
\end{figure}

More can be learned about the spin structure by inspecting the radial distribution of the spin up/down density $\rho_{\pm}(r)=\langle\Psi^\dag(\bm{r})[(1\pm\sigma_z)/2]\Psi(\bm{r})\rangle$. For $N=4,5,6,7$ (the cases $N=4$ and $N=7$ are shown in Fig.\ \ref{fig2} for $\alpha=2$) the spin-down component is dominant close to the center of the dot, while the spin-up component is dominant closer to the edge. Directly at the edge, however, the spin-resolved densities are forced to be equal by the boundary conditions. Qualitatively, this means that the spin-down density is stiffer towards interactions because it is concentrated closer to the center than the spin-up density.
This behavior is consistent with the increase of the spin polarization $\left<S_z\right>$ when the interaction strength is increased: spin-up electrons can respond faster to electron-electron interactions so that it is energetically favorable to polarize the system.

\begin{figure}
	\centering
	\includegraphics[width=\linewidth]{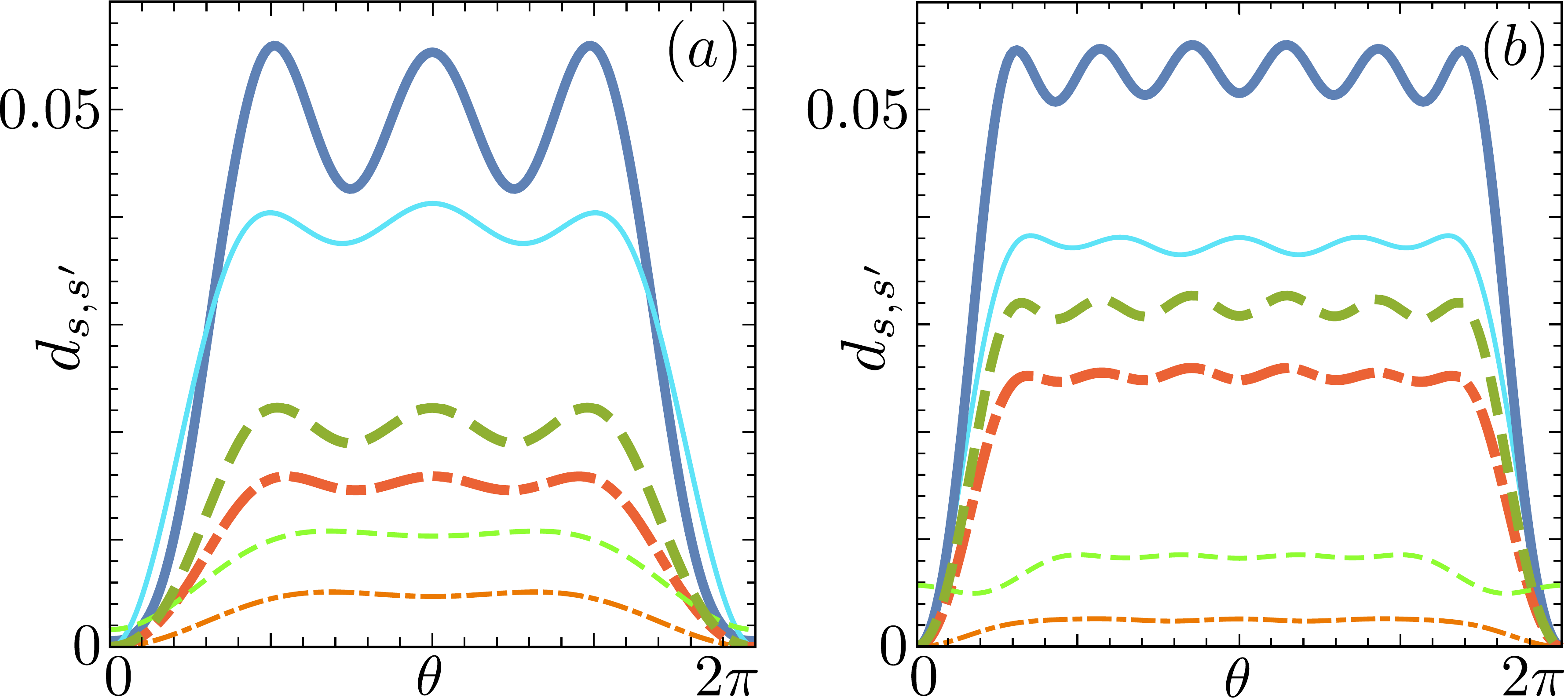}
	\caption{(Color online) Plot of $d_{+,+}(\bm{r},\bm{r}')$ (blue, solid), $d_{+,-}(\bm{r},\bm{r}')$ (green, dashed), and $d_{-,-}(\bm{r},\bm{r}')$ (red, dash-dotted), as a function of $\theta$; light/thin lines correspond to $\alpha=0$ and dark/thick lines to $\alpha=2$. In $(a)$ $N=4$, $\theta'=0$, and for $\alpha=0$, $r=r'=0.8R$, while for $\alpha=2$, $r=r'=0.93R$. (b) $N=7$, $\theta'=0$, and for $\alpha=0$, $r=r'=0.76R$, while for $\alpha=2$, $r=r'=0.96R$.}
	\label{fig3}
\end{figure}

In order to directly access the spin selection, we address the spin-resolved density-density correlation functions $d_{s,s'}(\bm{r},\bm{r}')$, with $s,s'=\pm$ labeling the spin projection. Explicitly, we define
\begin{equation}
d_{s,s'}(\bm{r},\bm{r}')=\frac{\langle\Psi^\dag_s(\bm{r})\Psi^\dag_{s'}(\bm{r}')\Psi_{s'}(\bm{r}')\Psi_s(\bm{r})\rangle}{N(N-\delta_{s,s'})},
\end{equation}
so that the interpretation of $d_{s,s'}(\bm{r},\bm{r}')$ is the probability of finding a particle with spin projection $s$ at position $\bm{r}$ if an electron with spin projection $s'$ is fixed at position $\bm{r}'$. The results for $N=4$ and $N=7$ (which well represent the physics up to seven particles) are shown in Fig.\ \ref{fig3} in the noninteracting case $\alpha=0$ and for $\alpha=2$. In the figure, $r$ and $r'$ are fixed at the maximum of the total density (spin-up density for $N=7$ and $\alpha=0$) and $d_{s,s'}(\bm{r},\bm{r}')$ is plotted as a function of the polar angle $\theta$ of $\bm r$ with $\theta'=0$. In the noninteracting case, we observe that $d_{+,+}(\bm{r},\bm{r}')$ dominates over the others due to the spin polarization induced by the boundary. Moreover, they are characterized by three and six peaks for respectively $N=4$ and $N=7$, which means that, on average, the weight of the spin-up component of each of the electrons is dominant. Consistent with the Pauli principle, $d_{+,+}(\bm{r},\bm{r}')$ and $d_{-,-}(\bm{r},\bm{r}')$ are zero for $\theta=0$, while $d_{+,-}(\bm{r},\bm{r}')$ is not. When interactions are increased the peaks in $d_{+,+}(\bm{r},\bm{r}')$ become more pronounced, indicating the onset of Wigner oscillations. The correlation function $d_{+,-}(\bm{r},\bm{r}')$ exhibits respectively three and six (weaker) oscillations as well. This is a consequence of the topological origin of the system. Remarkably, no signature of antiferromagnetic correlations, characterizing 2DEG dots \cite{antiferro}, are present. In our case, we also find $d_{+,-}(\bm{r},\bm{r}')\sim 0$ at $\theta=0$ which indicates unambiguously that $\alpha=2$ corresponds to significant interactions in the dot. The spin-down correlations $d_{-,-}(\bm{r},\bm{r}')$, however, remain suppressed (liquid-like) even for $\alpha=2$. The system is hence characterized by a spin selective Wigner oscillation. This behavior is consistent both with the increase in spin polarization and the radial dependence of the spin-resolved density. Intuitively speaking, one of the effects of interactions in confined systems is to push particles closer to the boundaries as to minimize the interaction energy. Particles that remain near the center when interactions are increased (the spin-down density in this case) do not minimize the interaction energy, and their spatial distribution and correlations are still mostly determined by kinetic energy and confinement. Therefore the system becomes more polarized with increasing interaction strength.

\begin{figure}
	\centering
	\includegraphics[width=0.94\linewidth]{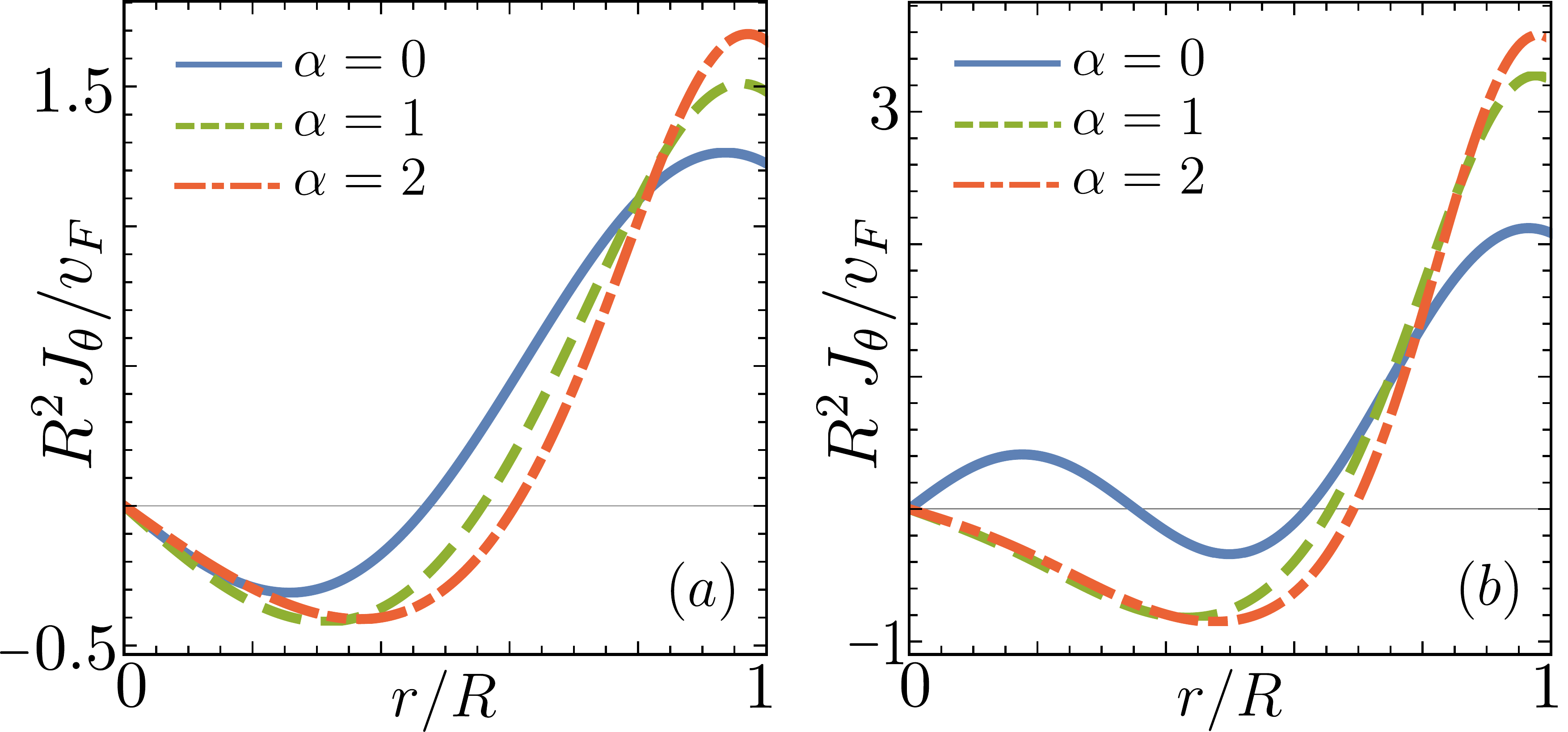}
	\caption{(Color online) Azimuthal current density $J_\theta(r)$ of the ground state for $\alpha=0$ (blue, solid), $\alpha=1$ (green, dashed), and $\alpha=2$ (red, dotted) for $(a)$ $N=4$, $(b)$ $N=7$.}
	\label{fig4}
\end{figure}

In the last part of this Letter, we discuss another consequence of the topological origin of the quantum dot on the local observables by inspecting the local azimuthal current density,
\begin{equation}
J_\theta(r)=v_F \hat \theta \cdot \langle \Psi^\dagger(\bm r) \bm \sigma \Psi(\bm r) \rangle,
\end{equation}
where $\hat\theta=(-y/r,x/r)$. Note that $J_\theta(r)$, unlike the quantities that we inspected until now, is not a spin-resolved observable, but depends crucially on the distribution of orbital angular momenta which are related to the spin degree of freedom by spin-momentum locking. An anomalous behavior of $J_\theta(r)$ can thus be expected in our system. The total angular momentum of the ground state for $\alpha=2$ is $J_z=2,4.5,4,7.5,12,10.5$ for respectively $N=2,..,7$ electrons. Consequently the ground state is a rotating Wigner oscillation with a nonzero local current. However, due to spin-momentum locking, the orbital angular momentum is not a good quantum number and the rotation pattern is nontrivial. As shown in Fig.\ \ref{fig4} for the cases (a) $N=4$ and (b) $N=7$, $J_\theta(r)$ changes sign as a function of $r$ with positive currents close to the edge and negative currents closer to the center. This behavior is due to contributions of negative angular momentum to the ground state, and goes beyond the intuitive picture of a coherently rotating Wigner oscillation for ordinary 2DEG quantum dots. Moreover, the boundary between regions of opposite current becomes sharper with increasing interaction strength. 

As far as the experimental observations of our predictions are concerned, different issues might arise. On the one hand, the realization of quantum dots on the surface of topological insulators remains a challenge but it should, for instance, be possible by magnetic confinement. On the other hand, the spin polarization that we predict, can, in principle, be accessed by coupling the dot to spin-polarized local probes \cite{spinpol1,spinpol2}. 
The texture characterizing $J_\phi(r)$ can be detected with nitrogen-vacancy centers in diamonds or nanoSQUIDs \cite{nv1,nv2,nanosquid}. In fact, for a dot with radius $R = 500$ nm, a classical finite element calculation shows that the typical variations of the resulting magnetic field are of the order of 0.2 $\mu T$, which is measurable by both techniques. In smaller dots, the signal, which scales as $1/R^2$, is larger, but, in that case, the required spatial resolution limits the use of nanoSQUIDs.

In conclusion, we have analyzed the ground-state properties of a quantum dot defined on the surface of a three-dimensional strong topological insulator. We have considered an intermediate interaction regime characterized by the onset of Wigner oscillations in the density-density correlation functions. The onset of Wigner oscillations is special in our system: It is concomitant with spin polarization of the dot, so that electron-electron interactions in the topological dot play a similar role as an external magnetic field for quantum dot in a 2DEG, and affects mainly one spin component.
Finally, we have discussed the behavior of the azimuthal particle current which changes sign as a function of the distance from the center of the dot, revealing a fingerprint of spin-momentum locking and hence of the topological origin of the system.

\begin{acknowledgments}
C.D.B.\ is an aspirant and M.Z.\ a postdoc fellow of the Flemish Research Foundation (FWO). N.T.Z.\ and B.T.\ acknowledge financial support by the DFG (SPP1666 and SFB1170 "ToCoTronics"), the Helmholtz Foundation (VITI), and the ENB Graduate school on "Topological Insulators". We thank F.\ Cavaliere, F.\ Cr\'epin, C.\ Felser, and B.\ Yan for interesting discussions, and S.\ Curreli for performing the finite element calculation of the magnetic field in COMSOL.
\end{acknowledgments}

\end{document}